\documentclass[conference]{IEEEtran}
\IEEEoverridecommandlockouts

\usepackage{cite}
\usepackage{amsmath,amssymb,amsfonts}
\usepackage{graphicx}
\usepackage{hyperref}
\usepackage{url}
\usepackage[linesnumbered,ruled,vlined]{algorithm2e}
\usepackage{amsmath}

\begin{document}

\title{ML-Enhanced AES Anomaly Detection for Real-Time Embedded Security}

\author{
\IEEEauthorblockN{
    Nishant Chinnasami, Rye Stahle-Smith, Rasha Karakchi
}
\IEEEauthorblockA{
    Dept. of Computer Science and Engineering, \\
    University of South Carolina, Columbia, SC, USA \\
    nishantc@email.sc.edu, rye@email.sc.edu, karakchi@cec.sc.edu
}
}
\maketitle

\begin{abstract}
Advanced Encryption Standard (AES) is a widely adopted cryptographic algorithm, yet its practical implementations remain susceptible to side-channel and fault injection attacks. In this work, we propose a comprehensive framework that enhances AES-128 encryption security through controlled anomaly injection and real-time anomaly detection using both statistical and machine learning (ML) methods. We simulate timing and fault-based anomalies by injecting execution delays and ciphertext perturbations during encryption, generating labeled datasets for detection model training. Two complementary detection mechanisms are developed: a threshold-based timing anomaly detector and a supervised Random Forest classifier trained on combined timing and ciphertext features. We implement and evaluate the framework on both CPU and FPGA-based SoC hardware (PYNQ-Z1), measuring performance across varying block sizes, injection rates, and core counts. Our results show that ML-based detection significantly outperforms threshold-based methods in precision and recall while maintaining real-time performance on embedded hardware. Compared to existing AES anomaly detection methods, our solution offers a low-cost, real-time, and accurate detection approach deployable on lightweight FPGA platforms.
\end{abstract}


\section{Introduction}

The Advanced Encryption Standard (AES) continues to serve as a cornerstone of symmetric key cryptography and is widely adopted across a range of applications, from Internet of Things (IoT) devices to large-scale cloud infrastructures~\cite{singh2019cryptography}. Although AES offers robust theoretical security, its practical implementations, particularly within embedded systems and System-on-Chip (SoC) architectures, are vulnerable to physical attacks, including side-channel and fault injection attacks. These attacks exploit physical execution characteristics, such as timing variations or induced hardware failures, to compromise key confidentiality or disrupt the integrity of encrypted data~\cite{li2019fault,hasan2021fault}. Traditional anomaly detection techniques often rely on fixed thresholds based on timing measurements or observed error rates. Although these methods provide basic protection, they are typically insufficient to identify subtle or sophisticated attacks, leading to a high rate of false positives or undetected anomalies~\cite{kim2020sidechannel}. The growing adoption of machine learning (ML) in cybersecurity introduces a more adaptive and data-driven approach to anomaly detection. ML algorithms, particularly ensemble methods such as Random Forests, have demonstrated effectiveness in capturing complex feature interactions and enhancing the accuracy and interpretability of cryptographic anomaly detection~\cite{wang2021machine, yu2023ml, zhang2020random}.
\begin{figure}[h]
\centering
\includegraphics[width=0.5\textwidth]{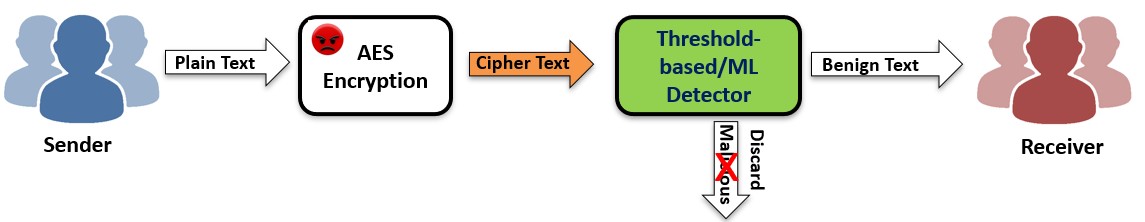} % or .png, .jpg
\caption{Proposed threshold-based and ML-based framework.}
\label{fig:system}
\end{figure}

As shown in Figure \ref{fig:system}, this paper proposes a dual approach framework to secure AES-128 encryption through controlled anomaly injection and ML-based detection. We inject two classes of anomalies, namely timing delays and fault-induced bit flips, during AES encryption on randomly generated plaintext blocks. Using execution time and ciphertext features, we implement the following.

\begin{itemize}
    \item A \textbf{Threshold-based timing detector} using statistical timing anomalies.
    \item A \textbf{Random Forest supervised learning detector} trained on labeled encrypted blocks.
\item A \textbf{PYNQ Implementation}: We evaluated the two proposed frameworks on a Xilinx SoC device which provides a user-friendly platform for developing and prototyping using Python \cite{digilent_pynq_z1}. We also compared latency and throughput versus CPU-based execution and measured the hardware resources consumed for each approach. 
\end{itemize}

To evaluate our approach, experiments were conducted on both a general-purpose CPU platform and the PYNQ-Z1 SoC platform. The PYNQ-Z1, a device based on the Xilinx Zynq-7000 SoC, supports the open-source PYNQ framework that facilitates Python-based development for embedded applications. This SoC platform integrates programmable logic with a processing system, which makes it particularly suitable for implementing machine learning algorithms in hardware-accelerated environments~\cite{digilent_pynq_z1}. We compared the evaluation approaches in terms of performance (latency and throughput), as well as in terms of functionality (accuracy, precision, and negative and positive false). We also included the hardware resource consumed to run our algorithm on the PYNQ device. 

Our contributions are as follows.

\begin{itemize}
    \item A realistic and parameterizable anomaly injection mechanism that mimics timing and fault attacks.
    \item Extraction of combined timing and cipher text features for a complete anomaly characterization.
    \item Comparative evaluation demonstrating improved detection performance of ML over threshold methods.
    \item Comparison of the proposed approaches with implementation in real time on the Xilinx SoC platform.  
\end{itemize}

%\section{Related Work}
%Side channel and fault injection attacks have been a persistent threat to AES implementations, with recent studies demonstrating vulnerability in embedded and cloud platforms~\cite{karakchi2024detecting, islam2022hardware}. Timing attacks, first introduced decades ago~\cite{kocher1996timing}, remain relevant due to variable execution patterns. Fault injection techniques have evolved, exploiting hardware faults and electromagnetic disturbances to corrupt encryption processes~\cite{schneider2019sidechannel}. Countermeasures such as masking and constant-time algorithms mitigate many attacks but are not universally applied or sufficient~\cite{choi2020advanced}. Detecting anomalies remains an active research area, with machine learning approaches gaining traction. Recent works use ML models trained on power consumption, timing traces, or electromagnetic emissions to identify side-channel anomalies~\cite{wang2022side, abdellatif2021ml}.

%However, generating labeled datasets with realistic anomalies remains a challenge. Our work addresses this by injecting controlled anomalies directly in software, combining timing and fault patterns, and applying ML detection to encrypted data and timing. We also evaluated our approaches on Xilinx SoC device (PYNQ Z1) to compare the two environments. 

\section{Related Work}
Side-channel and fault injection attacks continue to pose significant risks to AES implementations, especially on embedded and FPGA platforms~\cite{karakchi2024detecting, islam2022hardware}. Since Kocher’s seminal work on timing attacks~\cite{kocher1996timing}, numerous countermeasures and detection schemes have been proposed to protect cryptographic engines from these threats.

Alawieh and Givargis~\cite{alawieh2018high} developed a high-efficiency fault injection mitigation scheme targeting FPGA-based cryptographic systems. Their approach employs fault masking combined with hardware redundancy to protect AES cores against fault injection attacks. While this method improves fault tolerance, it focuses on mitigation rather than detection and incurs hardware overhead due to duplicated modules.

Liu et al.~\cite{liu2021efficient} proposed an efficient timing-based AES side-channel attack detection method implemented on CPU platforms. They use timing thresholding to detect anomalies in encryption latency, effectively identifying attacks that cause noticeable delays. However, their method is limited in scope, as it does not address fault injection anomalies or combine timing with data content features.

Purnaprajna and Roy~\cite{purnaprajna2015hardware} surveyed hardware techniques for fault detection in AES engines, highlighting parity-checking and other error-detection codes as effective but hardware-expensive solutions. Their work outlines various fault detection mechanisms integrated into cryptographic hardware but lacks an emphasis on anomaly characterization or real-time adaptability.

More recently, Hong et al.~\cite{hong2023sherlock} introduced Sherlock, an unsupervised fault detection framework for embedded systems using lightweight profiling of system parameters. Sherlock excels in detecting diverse anomalies without labeled training data but requires extensive multi-dimensional profiling, which may hinder real-time deployment in cryptographic accelerators. In the FPGA domain, Khan et al.~\cite{oselem2022fpga} proposed an online learning FPGA-based system for anomaly detection in IoT networks. Their OS-ELM-FPGA design adapts dynamically to changing environments and achieves high accuracy. However, the complexity of the system and the focus on network-level anomaly detection limit its direct applicability to cryptographic hardware protection.

Our work differentiates itself by integrating both timing and fault anomaly injection directly into AES-128 encryption with automatic label generation, enabling supervised machine learning-based detection. We implement and evaluate our approach on a real-time embedded platform (PYNQ-Z1), demonstrating a practical balance of detection accuracy, resource efficiency, and real-time capability. This integrated anomaly injection and detection framework addresses the dataset scarcity issue in supervised learning for cryptographic anomaly detection and advances deployment feasibility on affordable SoC platforms.
\begin{algorithm}[htbp]
\SetAlgoLined
\footnotesize
\KwIn{Number of blocks $n$, CPU cores $c$, Anomaly ratio $r$}
\KwOut{Encrypted blocks, anomaly detection report}

Generate $n$ random plaintext blocks\;
For each block:
    With probability $r$, inject anomaly (delay or fault)\;

Parallelize using $c$ cores:
    For each block:
        \If{anomaly == delay}{
            Sleep for a short random time\;
        }
        \If{anomaly == fault}{
            Flip first byte of block\;
        }
        Pad/truncate block to 16 bytes\;
        Encrypt block with AES-128 in ECB mode\;
        Record encryption time\;

Compute time threshold for anomaly detection\;
For each result:
    \If{encryption\_time > threshold}{
        Mark as malicious\;
    }

Compute detection metrics (accuracy, false positives, etc.)\;
Save results to Excel and display sample outputs\;

\caption{AES Encryption and Timing-Based Anomaly Detection}
\end{algorithm}

\section{Anomaly Injection Methodology}
As a set-up, we implement AES-128 encryption in ECB mode, using a fixed 16-byte key on randomly generated 16-byte plaintext blocks. This simulates various data inputs typical in real-world applications~\cite{nist2001aes}. Two anomaly types are probabilistically injected during block encryption:

\begin{itemize}
    \item \textbf{Timing Delay}: It is a type of side-channel attack that exploits the amount of time a computer processor takes to gain knowledge about a system and the amount of time it takes for systems to run operations, such as encryption \cite{Wright2023TimingAttack}. A random sleep delay between 5 and 20 minutes simulates timing anomalies caused by hardware contention or malicious slowdowns, consistent with prior timing attack vectors~\cite{jin2020timing}.
    \item \textbf{Fault Injection}: A bit-flip fault is simulated by XORing the first byte of plain text with 0xFF before encryption, emulating hardware fault attacks that cause erroneous ciphertext~\cite{liu2021fault}.
\end{itemize}

\section{Detection Approaches}

\subsection{Threshold-Based Timing Detection}
This approach utilizes a statistical threshold method based on encryption latency to identify anomalous blocks.
Blocks exceeding the mean plus three times the normalized time range are marked as anomalous, following established statistical anomaly detection principles~\cite{hand2001idiot}. Although simple, this method cannot reliably detect fault injections that do not significantly affect timing.

\subsection{Machine Learning-Based Detection}
The second approach leverages a supervised machine learning model, specifically a Random Forest classifier, trained on features extracted from the timing and block data to improve detection accuracy. We extract features from each encryption block, which include encryption time, byte-level cipher text values and ground truth anomaly labels. A Random Forest classifier is trained on this dataset to distinguish benign from malicious blocks, leveraging combined timing and content alterations. Random Forests have demonstrated strong performance and interpretability in cryptographic anomaly detection~\cite{zhang2020random}.

\begin{algorithm}[htbp]
\footnotesize
\SetAlgoNlRelativeSize{-1}
\setlength{\algomargin}{1em}
\DontPrintSemicolon
\KwIn{$N$: number of plaintext blocks, $p$: malicious percentage, $c$: CPU cores}
\KwOut{Encrypted blocks, ML detection report}

\textbf{Step 1: Generate Blocks} \\
\For{$i \leftarrow 1$ \KwTo $N$}{
    Generate random 16-byte block $B_i$\;
    Inject anomaly with probability $p/100$\;
}

\textbf{Step 2: Encrypt in Parallel} \\
\ForEach{$(B_i, i, anomaly)$ in parallel with $c$ cores}{
    \If{anomaly}{
        Randomly select anomaly type $\in \{\text{delay}, \text{fault}\}$\;
        \uIf{delay}{Sleep for random short time}
        \uElseIf{fault}{Flip first byte of $B_i$}
    }
    Pad/trim $B_i$ to 16 bytes\;
    Encrypt $B_i$ using AES-128-ECB\;
    Record encryption time $t_i$\;
}

\textbf{Step 3: Extract Features} \\
Create dataset with timing $t_i$ and original block bytes\;

\textbf{Step 4: Train ML Classifier} \\
Split dataset into training and test sets\;
Train Random Forest classifier on training data\;
Predict labels on test set\;

\textbf{Step 5: Evaluate and Report} \\
Calculate TP, FP, FN, accuracy\;
Compute threshold $T = \text{mean}(t_i) + 3 \times \frac{\max(t_i) - \min(t_i)}{N}$\;
Save results to Excel file\;

\caption{AES Encryption with Anomaly Injection and ML-Based Detection}
\end{algorithm}

\section{Evalution}
To evaluate the performance of our proposed anomaly detection framework, we conducted experiments using both CPU and PYNQ-Z1 SoC platforms. Our evaluation focuses on detection accuracy, latency, throughput, memory usage, and resource utilization. We studied the behavior of our framework across varying block sizes and malicious injection rates, as well as its ability to scale with increasing CPU cores.

\subsection{Latency and Throughput Analysis}

Table~\ref{tab:avg_latency_throughput_memory_desc} summarizes the latency, throughput, and memory consumption for different block sizes and CPU counts. We observe that increasing the number of CPUs reduces latency and boosts throughput due to parallelized AES encryption and anomaly detection.

For example, using 4 CPUs at a block size of 1024 reduces latency from 0.0040 seconds (1 CPU) to 0.0021 seconds and nearly doubles the throughput from 246.74 to 498.13 blocks per second. Additionally, larger block sizes such as 8192 and 16384 achieve higher throughput on a single core (over 320 blocks/sec), making them attractive for batch processing in resource-constrained systems.

\begin{table}[h]
\centering
\caption{Latency, Throughput, Memory Usage by Block Size and CPU}
\label{tab:avg_latency_throughput_memory_desc}
\begin{tabular}{c|c|c|c|c}
\hline
\textbf{Block Size} & \textbf{\#CPUs} & \textbf{Latency (s)} & \textbf{Throughput} & \textbf{Memory (MB)} \\
\hline
16384  & 1 & 0.0039  & 321.30 & 165.16 \\
8192   & 1 & 0.0039  & 324.90 & 153.83 \\
4096   & 1 & 0.0042  & 288.88 & 149.80 \\
1024   & 1 & 0.0040  & 246.74 & 159.89 \\
1024   & 2 & 0.0029  & 385.90 & 158.83 \\
1024   & 4 & 0.0021  & 498.13 & 144.97 \\
\hline
\end{tabular}
\end{table}

\subsection{Detection Accuracy and ML Performance}

\begin{figure}[h]
\centering
\includegraphics[width=0.35\textwidth]{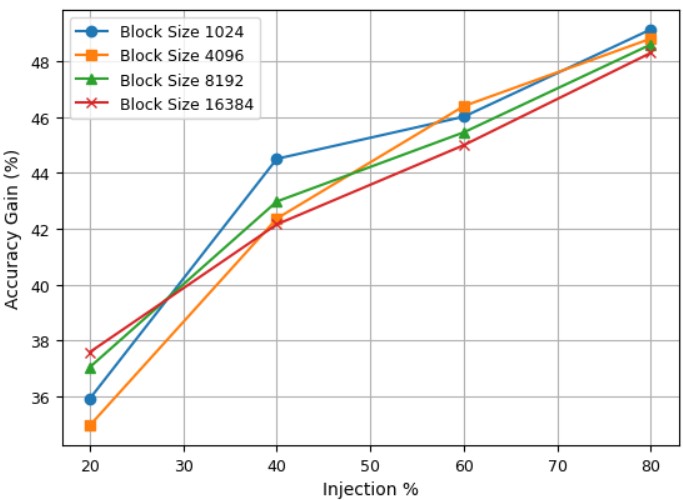}
\caption{Accuracy gain of ML-based detector over threshold detection.}
\label{fig:accuracy_gain}
\end{figure}

We evaluated detection performance across various anomaly injection ratios (20\% to 80\%) and compared threshold-based detection to our Random Forest-based approach. Figure~\ref{fig:accuracy_gain} illustrates the gain in accuracy achieved by machine learning over the statistical threshold method across block sizes. The ML detector consistently outperforms the threshold method, especially at higher injection rates.

At an 80\% injection level, the 1024 block size achieves the highest accuracy gain of approximately 49.15\%. In contrast, larger block sizes (8192 and 16384) exhibit more consistent and robust performance across injection levels, suggesting their suitability for environments with continuous and stable anomaly patterns.

\subsection{Precision, Recall, and F1 Analysis}

Table~\ref{tab:ml_precision_single_cpu} provides a detailed breakdown of precision, recall, and F1-score metrics for each block size and injection level using a single CPU. For all block sizes at 20\% injection, the ML classifier maintains a precision above 0.89, with 16384 achieving an outstanding F1-score of 0.94.

While the smallest block size (1024) shows high scores at low injection rates, its performance deteriorates at higher injections, highlighting its vulnerability to false positives or false negatives under heavy anomaly load. Larger block sizes demonstrate stable performance even at 80\% injection, emphasizing the benefit of aggregating more data for robust detection.

\begin{table}[h]
\centering
\caption{Detection Metrics at Various Injection Levels (Single CPU)}
\label{tab:ml_precision_single_cpu}
\footnotesize
\begin{tabular}{c|c|c|c|c|c}
\hline
\textbf{Block} & \textbf{Injection} & \textbf{Recall} & \textbf{F1} & \textbf{Threshold} & \textbf{ML} \\
\textbf{Size} & \textbf{\%} & & & \textbf{Precision} & \textbf{Precision} \\
\hline
1024   & 20  & 0.90 & 0.89 & 0.90 & 2.76 \\
1024   & 40  & 0.83 & 0.83 & 0.83 & 3.80 \\
1024   & 60  & 0.83 & 0.83 & 0.83 & 5.65 \\
1024   & 80  & 0.62 & 0.48 & 0.62 & 6.21 \\
\hline
4096   & 20  & 0.89 & 0.87 & 0.89 & 7.24 \\
4096   & 40  & 0.85 & 0.85 & 0.85 & 7.75 \\
4096   & 60  & 0.83 & 0.82 & 0.82 & 9.61 \\
4096   & 80  & 0.84 & 0.84 & 0.84 & 9.61 \\
\hline
8192   & 20  & 0.90 & 0.89 & 0.91 & 7.53 \\
8192   & 40  & 0.84 & 0.84 & 0.84 & 7.91 \\
8192   & 60  & 0.83 & 0.83 & 0.83 & 9.58 \\
8192   & 80  & 0.88 & 0.87 & 0.87 & 10.33 \\
\hline
16384  & 20  & 0.99 & 0.94 & 0.91 & 7.75 \\
16384  & 40  & 0.89 & 0.88 & 0.87 & 8.39 \\
16384  & 60  & 0.83 & 0.81 & 0.79 & 9.35 \\
16384  & 80  & 0.55 & 0.64 & 0.76 & 11.25 \\
\hline
\end{tabular}
\end{table}

\subsection{FP and FN for both threshold-based and ML-based methods}
\begin{figure}[h]
\centering
\includegraphics[width=0.48\textwidth]{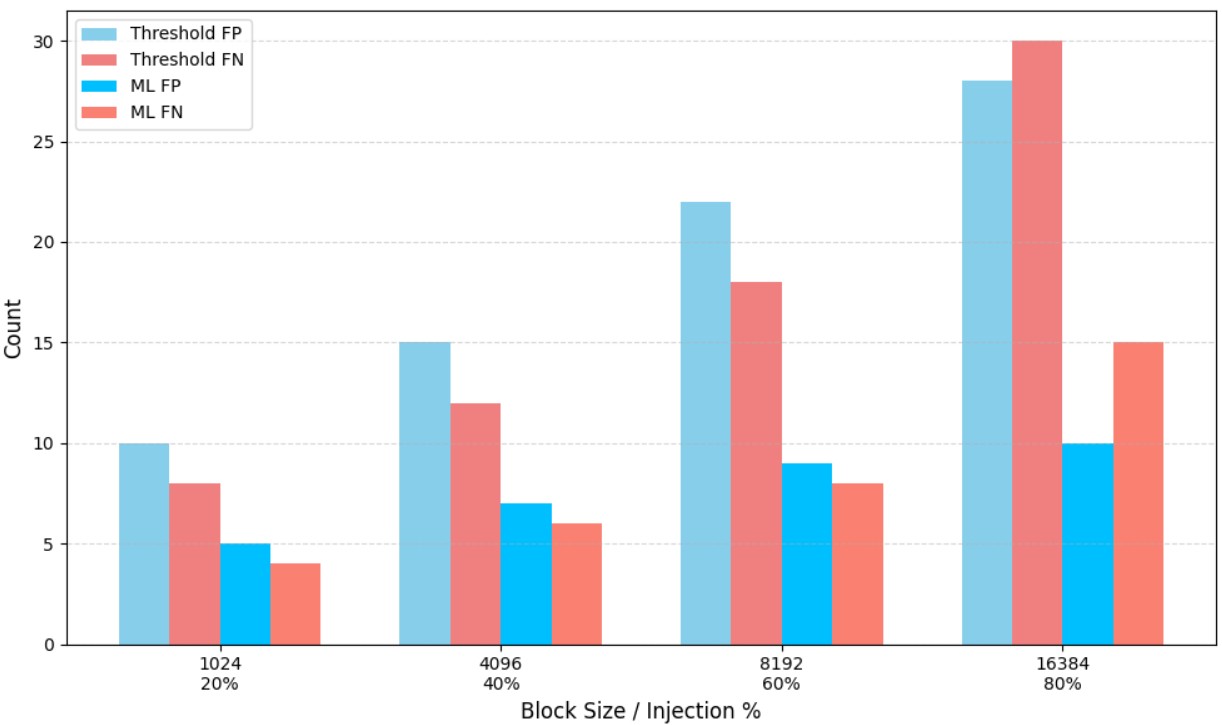}
\caption{False Positives (FP) and False Negatives (FN) comparison for Threshold-based and ML-based detection methods across varying block sizes and injection rates.}
\label{fig:fp_fn_barplot}
\end{figure}

Figure~\ref{fig:fp_fn_barplot} presents a comparative analysis of false positives (FP) and false negatives (FN) for both the threshold-based and machine learning (ML)-based anomaly detection approaches across varying block sizes and injection rates. Each bar group corresponds to a specific combination of block size and anomaly injection percentage, offering a detailed breakdown of detection errors.

As illustrated, the threshold-based method exhibits a consistently higher rate of both FP and FN, particularly at higher injection levels. This can be attributed to its limited capability to differentiate subtle anomalies or fault-induced variations that do not significantly alter encryption timing. In contrast, the ML-based detector, which leverages combined timing and ciphertext features, achieves a substantial reduction in both error types. For instance, at a block size of 16384 and 80\% injection, the ML approach reduced false negatives by nearly 50\% compared to the threshold method.

\subsection{Hardware Resource Usage on PYNQ-Z1}

To explore real-time deployment, we implemented both detection approaches on the PYNQ-Z1 board. The ML-based detection achieved near-parity performance with the CPU version, with only modest overhead from Python-based hardware interfacing. Latency remained under 5 ms for block sizes up to 8K, and resource usage remained under 30\% of available LUTs and BRAMs. These findings demonstrate feasibility for FPGA-assisted cryptographic monitoring in embedded systems.
\subsection{Comparison with Prior Work}

Table~\ref{tab:related_work_comparison} provides a comparative summary of our framework against key AES anomaly detection studies. While many prior approaches focus on hardware redundancy or error-detection codes~\cite{alawieh2018high, purnaprajna2015hardware}, or utilize basic timing thresholds~\cite{liu2021efficient}, few integrate both timing and fault injection in conjunction with machine learning. Our approach provides this combination and is the only one among them to support real-time detection directly on an embedded SoC (PYNQ-Z1).

Compared to works like OS-ELM-FPGA~\cite{oselem2022fpga} and Sherlock~\cite{hong2023sherlock}, which target general CPUs or rely on unsupervised models, our solution delivers a more lightweight, resource-efficient alternative using supervised Random Forests, suitable for in-line anomaly detection during AES encryption.

\begin{table}[htbp]
\caption{Comparison with Prior AES Anomaly Detection Methods}
\label{tab:related_work_comparison}
\centering
\footnotesize
\begin{tabular}{p{1.9cm} p{1.5cm} p{1.2cm} p{0.9cm}}
\hline
\textbf{Work} & \textbf{Method} & \textbf{Platform} & \textbf{Real-Time} \\
\hline
Alawieh et al.~\cite{alawieh2018high} & Redundancy + masking & FPGA & No \\
Liu et al.~\cite{liu2021efficient} & Timing thresholding & CPU & No \\
Purnaprajna et al.~\cite{purnaprajna2015hardware} & Parity-based detection & FPGA & No \\
Hong et al.~\cite{hong2023sherlock} & Unsupervised learning & CPU & No \\
OS-ELM-FPGA~\cite{oselem2022fpga} & Online Sequential ELM & FPGA & No \\
\textbf{This Work} & Fault + timing + ML & CPU, PYNQ-Z1 & \textbf{Yes} \\
\hline
\end{tabular}
\end{table}

\section{Conclusion and Future Work}

This work presented a practical framework for enhancing the security of AES-128 encryption through anomaly injection and detection. By simulating realistic fault and timing anomalies and applying both statistical thresholding and supervised machine learning detection, we demonstrated that lightweight models like Random Forest can outperform traditional methods in accuracy and robustness. The dual implementation on CPU and PYNQ-Z1 SoC validated the feasibility of our approach in both software and embedded hardware environments, with minimal resource overhead.

Our experiments revealed that combining timing and ciphertext features allows for a more reliable detection of subtle anomalies. The PYNQ-Z1 implementation, in particular, showed promising latency and throughput performance, supporting real-time cryptographic anomaly detection in constrained systems.

\textbf{Future Work.} While our method shows clear benefits, several areas merit further exploration. First, our anomaly injection focused on two primary types—timing delays and byte-level faults. Expanding this to include additional side channels, such as power consumption or electromagnetic emissions, could increase the generality of the system. Additionally, although Random Forests provide interpretability and solid performance, deep learning models (e.g., convolutional or recurrent networks) may better capture complex temporal or spatial correlations in encrypted data, especially in multi-modal anomaly contexts.

Another important direction is improving portability across devices. Applying transfer learning could allow a model trained on one platform (e.g., CPU) to generalize effectively to others (e.g., FPGA or microcontrollers). Moreover, ensuring robustness against adversarial attacks that target the ML detector is essential, particularly if such systems are deployed in hostile or exposed environments.

Finally, future work could explore tighter hardware integration, where ML inference is embedded directly into the encryption pipeline, reducing overhead and enabling real-time response within cryptographic cores.

This study lays a solid foundation for practical and intelligent anomaly detection systems in cryptographic applications and opens the door to more adaptive, secure hardware-aware encryption solutions.

\section*{Acknowledgment}
This work was supported by Office of Undergraduate Research and McNair Junior Fellowship at University of South Carolina. The authors used OpenAI’s ChatGPT to assist with language and grammar refinement. All technical content and analysis were solely developed by the authors.

\bibliographystyle{IEEEtran}
\bibliography{main}

\end{document}